\documentclass[utf8]{frontiersSCNS} 

\usepackage{url,hyperref,lineno,microtype,subcaption}
\usepackage[onehalfspacing]{setspace}
\usepackage{tablefootnote}

\def\Msun{$M_{\odot}$}

\def\keyFont{\fontsize{8}{11}\helveticabold }
\def\firstAuthorLast{Zwintz {et~al.}} 
\def\Authors{Konstanze Zwintz\,$^{1,*}$ and Thomas Steindl\,$^{1}$}
\def\Address{$^{1}$Institute for Astro- and Particle Physics, University of Innsbruck, Innsbruck, Austria}
\def\corrAuthor{Konstanze Zwintz}

\def\corrEmail{konstanze.zwintz@uibk.ac.at}

\begin{document}

\onecolumn
\firstpage{1}

\title[Pre-MS Asteroseismology]{The Pre-main Sequence: Challenges and Prospects for Asteroseismology} 

\author[\firstAuthorLast ]{\Authors} 
\address{} 
\correspondance{} 

\extraAuth{}

\maketitle

\begin{abstract}

\section{}
Stars do not simply pop up on the main sequence. Before the stars arrive on the zero-age main sequence, they form in the collapses of molecular clouds, gain matter through accretion processes, and compress their cores until hydrogen can burn in full equilibrium. Although this evolutionary phase lasts a relatively short time, it is the imprint of these important physical processes that is often ignored by simplified assumptions. While asteroseismology offers a great tool to investigate these physical processes, studying pre-MS oscillations in turn has the potential to further advance the field. 

Asteroseismology of pre-main sequence stars faces observational and theoretical challenges. The remnants of their birth environment which is often still surrounding the young stars causes variability that can interfere with the signal of pulsations. The lack of long time-base satellite observations in addition limits the applications of the method. Theoretical models of pre-main sequence stars include several assumptions and simplifications that influence the calculation of pulsation frequencies and excitation properties of pulsation modes.

Keeping all this in mind, the prospects for pre-main sequence asteroseismology are manifold. An improved understanding of the structure of young stellar objects has the potential to answer some of the open questions of stellar evolution, including angular momentum transport and the formation of magnetic fields. While gyrochronology, for example, struggles to determine the ages of the youngest clusters, pulsations in pre-main sequence stars can function as an independent age indicator yielding higher precision for single stars.

The increasing interest of stellar astrophysics in general to investigate the formation and early evolution of stars and planets illustrates the growing importance of pre-main sequence asteroseismology. In this work we discuss its potential for an advancement of our understanding of stellar structure and evolution.

\tiny
 \keyFont{ \section{Keywords:} early stellar evolution, pre-main sequence, p- and g-mode pulsations, stellar structure, accretion physics, angular momentum transport, asteroseismology} 
\end{abstract}

\section{Introduction}

The study of pre-main sequence stars was initiated in the 1950s when \textit{what appear to be recently formed groups of stars} \citep{Henyey1955} drew interest from the astronomical community. \citet{Henyey1955} provided the first calculations of stars before their main sequence phase. Their models described the gravitational contraction of radiative stars and the respective evolution of the spectroscopic parameters is still referred to as the `Henyey track' today. Once it was evident that convection plays a major part in the evolution of stars, \citet{Hayashi1961} delivered improved theoretical models for the pre-main sequence phase, achieving good agreement with the observational data of NGC 2264 \citep{Walker1956}. \citet{Hayashi1961} discussed the forbidden zone in the Hertzsprung-Russell diagram -- an area in which no star can be in the hydrostatic equilibrium as the needed temperature gradient would immediately be brought down by rapid convection -- and provided calculation after which stars follow a fully convective `Hayashi track' before joining the `Henyey track' on their contraction towards the ZAMS. Because of the forbidden zone in the Hertzsprung-Russell diagram, the models by \citet{Hayashi1961} follow the `Hayashi track' before joining the `Henyey track' on their contraction towards the ZAMS. \citet{Iben1965} refined the picture of pre-main sequence evolution (classical pre-main sequence model from here on) by following the ${\rm C}^{12}$-depletion in more detail 

Compared to the real star formation process, however, this classical view of the pre-main sequence evolution suffers from a crude approximation: the initial model. While the latter is taken as a huge ($\sim 55\,R_\odot$ for a $2\, M_\odot$ star) fully convective star at ZAMS mass, real stellar seeds are produced in the collapse of molecular clouds. Such an optically thin cloud collapses under its own gravity. The increase in density and temperature leads to the formation of a first hydrostatic core which will further heat up until molecular hydrogen dissociates at $\sim2000$\,K. This is a strongly endothermic process and leads to a second collapse, ending in the formation of the second hydrostatic core \citep[see e.g.][]{Larson1969}. Such a stellar seed with $1-5\,R_\odot$ and $10^{-3}-10^{-2}\,M_\odot$ \citep{Larson1969, Bhandare2018}, constitutes the first stage of the pre-main sequence evolution, and continues to accrete material from its surrounding cloud or disk.

The evolution of such accreting protostars was followed by multiple authors including \citet{Palla1990}, who phrased the word `birthline' meant as the position in the Hertzsprung-Russell diagram in which the radius of the accreting protostar first coincides with the radius of the classical pre-main sequence models. This created a misconception: a view in which stars evolve along the classical pre-main sequence tracks but are still hidden underneath their dust cloud and become visible when they cross the birthline. This view is a very unphysical picture, as stars evolve along the birthline (or rather their very own track) during their accretion phase. The concept of such a birthline is hence outdated, with state-of-the-art models of the pre-main-sequence providing a very different picture. The latter has been manifested by many authors \citep[e.g.][]{Hartmann1996, Hartmann1997, Wuchterl2001, Baraffe2009, Baraffe2010, Hosokawa2011, Baraffe2012, Kunitomo2017, Jensen2018, Elbakyan2019}, but only recently arrived in the field of asteroseismology \citep{Steindl2021b, Steindl2022a, Steindl2022b}.

\begin{figure}[h!]
\begin{center}
\includegraphics[width=\linewidth]{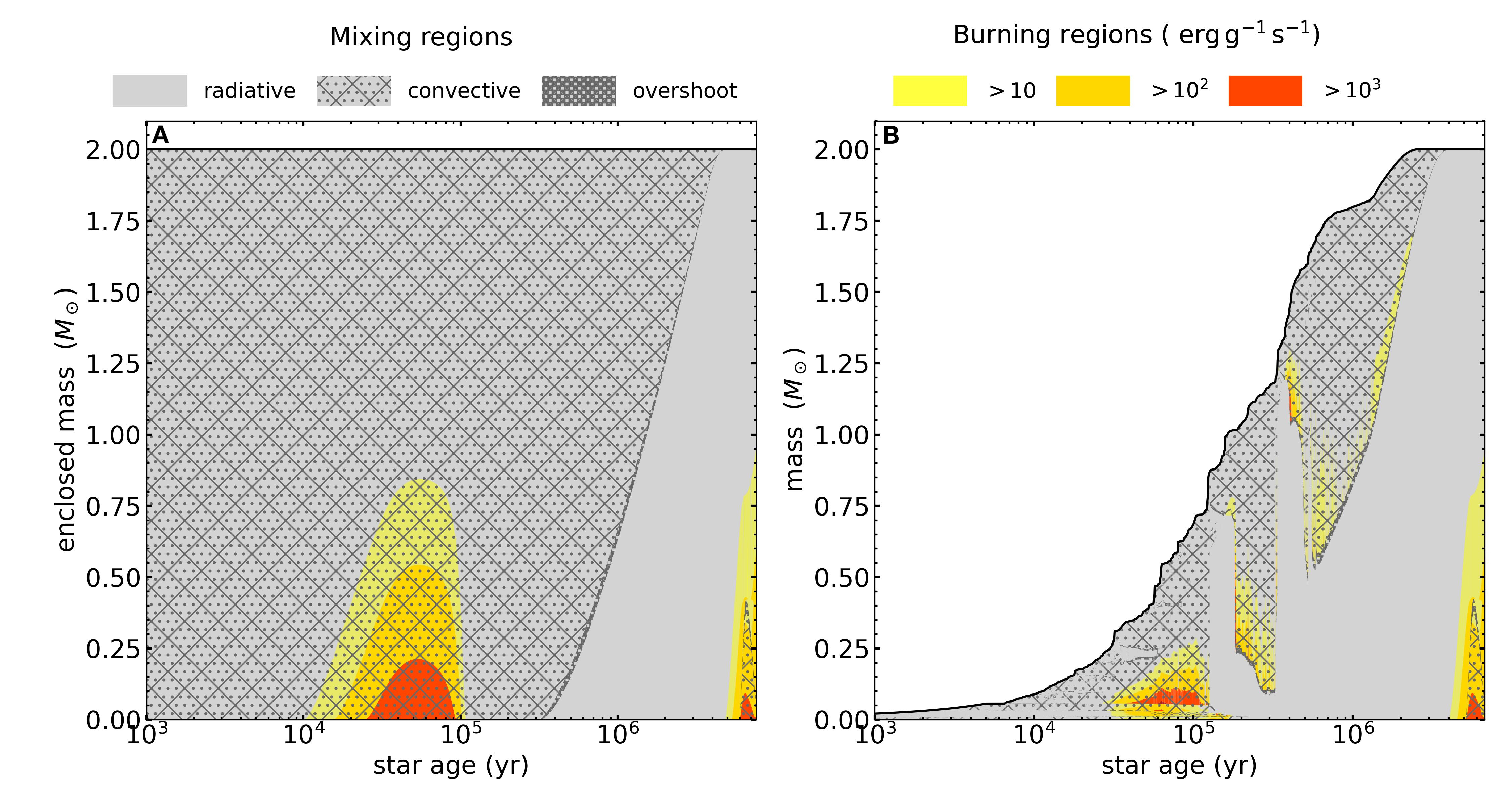}
\end{center}
\caption{Kippenhahn diagram of the pre-main sequence evolution. Left panel: classical model. Right panel: model including accretion effects \citep{Steindl2022a} with disk-mediated accretion rates from \citep{Elbakyan2019}. The stellar structure is shown between the stellar centre (enclosed mass = 0) and the black line (enclosed mass = stellar mass). Different mixing regions are indicated by textures. Radiative regions are grey, while overshooting and convective regions are hashed according to the legend. Regions of the star producing energy by nuclear reactions are colored according to the energy rate and the legend. }
\label{fig:kippenhahn}
\end{figure}

Introducing accretion effects into the numerical simulations of the pre-main sequence evolution provides an insight into the complicated structure of such young stars. The Kippenhahn diagram in Figure \ref{fig:kippenhahn} shows the striking differences between the simplified classical model (\ref{fig:kippenhahn}A) and the more realistic simulation including disk-mediated accretion rates (\ref{fig:kippenhahn}B). Most notable is the difference in chemical mixing. While the view of fully convective pre-main sequence stars is deeply rooted, state-of-the-art pre-main sequence models show that this is not the case although a large part of the stellar interior can still be affected by convection at different stages of the pre-main sequence evolution.

\begin{figure}[h!]
\begin{center}
\includegraphics[width=\linewidth]{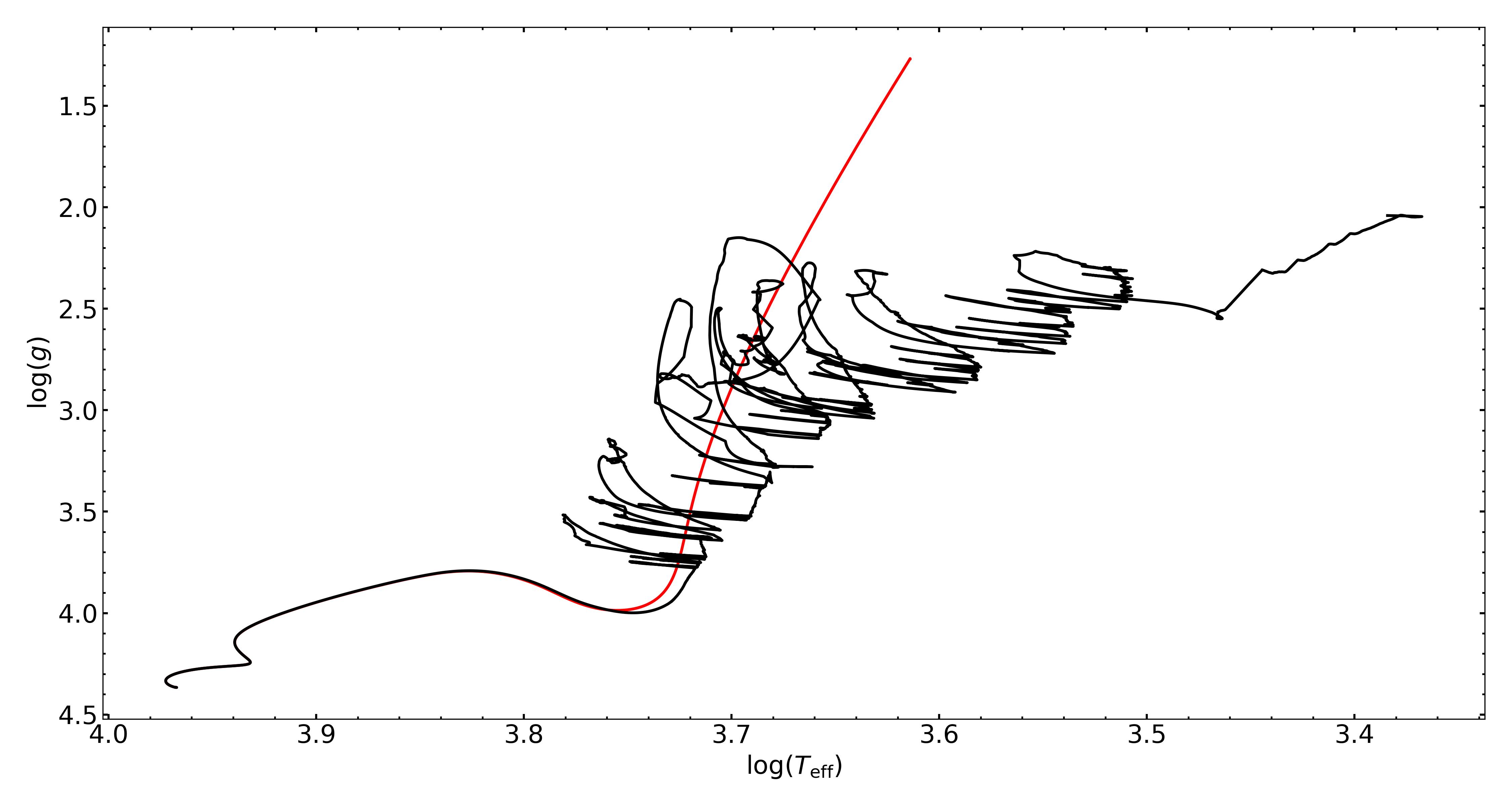}
\end{center}
\caption{Kiel diagram of the pre-main sequence models shown in Figure \ref{fig:kippenhahn}. The classical mode (red) follows the well known Hayashi track before entering the Henyey track and contracting on to the ZAMS. The disk mediated model (black) shows the effect of time-dependent accretion rates. The evolution follows a very chaotic route before entering the Henyey track and converging with the classical model.}
\label{fig:kiel}
\end{figure}

Naturally, such drastic changes in internal structure are also mirrored by the spectroscopic parameters of the star. Figure \ref{fig:kiel} provides the corresponding evolutionary tracks in the Kiel diagram (log($g$)-log($T_{\rm eff}$)-diagram). It is important to state that the track for the disk-mediated model is unique, that is a different accretion history will lead to significantly changed evolutionary track. In order to provide a realistic picture of pre-main sequence evolution, we have to move on from simplified views (Hayashi-track $\rightarrow$ Henyey-track $\rightarrow$ main sequence) and have to accept more complicated evolutionary paths: The spectroscopic parameters and internal structure in the early phases of a stars' lifetime are directly related to the properties of the mass accretion. Only after the disk has dissolved and the star continues to evolve without obtaining new material, will the structure of real stars gradually converge towards the structure that we are used to from the classical models. Even if spectroscopic parameters are rather similar, the internal structure remains different. Most notably is the existence of a temperature inversion towards the centre of the star \citep[see e.g. Fig. 8 of][]{Steindl2021b}. An imprint of star formation on the internal structure remains throughout the pre-main sequence phase and at least until the ZAMS; this should provide the opportunity to probe such disk-mediated evolution models with asteroseismology \citep{Steindl2022a}. 

Astrophysicists of the past have long desired a tool to probe the stellar interior. While direct photometric and spectroscopic methods pierce only the stellar atmosphere, information about the entire star are needed to improve our theory of stellar structure and evolution. Today, such a tool is available. Asteroseismology -- the theory of stellar oscillations -- provides the opportunity to measure (often) tiny changes in the stellar structure powered by stellar pulsations by means of photometric or spectroscopic methods \citep[overviews about asteroseismology can be found in, e.g.,][]{JCD1982,Gough1987,Unno1989,Aerts2010}. As the pulsations travel throughout the star, their frequencies hold information about the entire structure hence providing a view deep into the stellar interior. Since its discovery, asteroseismology has allowed many improvements of our understanding of stellar structure throughout the entire Hertzsprung-Russell diagram and all evolutionary stages \citep[e.g.,][]{Aerts2021}. 

Especially promising is the research field ``Pre-main sequence asteroseismology". Its origin lies in the first discovery of pulsations in young stars by \citet{Breger1972}. In his article, he reported that the two members of the young open cluster NGC\,2264 -- V\,588\,Mon and V\,589\,Mon -- show $\delta$ Scuti-type pulsations. But it took 20 years until the next observational detections were available \citep[e.g.,][]{Praderie1991,Kurtz1995}. These observations triggered the search for additional members of this new group of pulsating stars, the pre-main sequence $\delta$ Scuti stars, as well as the first theoretical work on pulsational instability in stars before the onset of hydrogen core burning \citep{Marconi1998}.
Subsequently, more pre-main sequence stars were found to show radial and non-radial oscillations. It soon became obvious that not only $\delta$ Scuti type pulsations can be excited in young stars, but also $\gamma$ Doradus \citep[e.g.,][]{Bouabid2011,Zwintz2013} and Slowly Pulsating B type variability \citep[e.g.,][]{Gruber2012}. A complete overview of the history of pre-main sequence asteroseismology can be found in \citet{Zwintz2019}.

A very important milestone in the field of pre-main sequence asteroseismology was the discovery of the presence of non-radial pulsations in pre-main sequence $\delta$ Scuti stars and the corresponding theoretical description \citep{Ruoppo2007,Zwintz2007}. Soon after, the observed pulsation frequencies of a $\delta$ Scuti star were used to confirm its pre-main sequence evolutionary stage in combination with theoretical models \citep{Guenther2007}. Pre-main sequence $\delta$ Scuti stars have since then proven to be a treasure trove for observational discoveries, with \citet{Zwintz2014} showing a connection between the pulsational properties of pre-main sequence $\delta$ Scuti stars and their relative evolutionary stage: the closer the stars are to the onset of hydrogen core burning, the faster they oscillate. Such a direct connection between stellar pulsation frequencies and the relative evolutionary stages has not yet been found for more evolved  $\delta$ Scuti stars.

More recent milestones include the discovery of a first candidate of solar-like oscillations in pre-main sequence stars by \citet{Muellner2021}, after predictions of their existence were already made early on by \citep{Samadi2005}. Furthermore, the case of RS Cha, a pre-main sequence eclipsing binary consisting of two $\delta$ Scuti stars, provides the best evidence for the discovery of tidally perturbed pulsations in young stars to date \citep{Steindl2021a}. Pre-main sequence asteroseismology provides the opportunity for many more exciting discoveries. To get there, however, many challenges have to be overcome, in order to uncover the mysteries of this complicated evolutionary stage. The aim of this work is to present these challenges and provide the reader with an outlook on the great prospects this field offers. We review pulsations in young stars in Section \ref{sec:puls_young_stars} before an in-depth description of the challenges pre-main sequence asteroseismology is faced with, both observational and theoretical, in Section \ref{sec:challenges}. An idea for a space mission dedicated to young stars and star forming regions is presented in Section \ref{STRETTO}. Section \ref{sec:prospects} concludes this work with a discussion of possible future milestones and how we might be able to achieve them sooner rather than later.

\section{Pulsations in young stars}
\label{sec:puls_young_stars}

\begin{figure}[h!]
\begin{center}
\includegraphics[width=\linewidth]{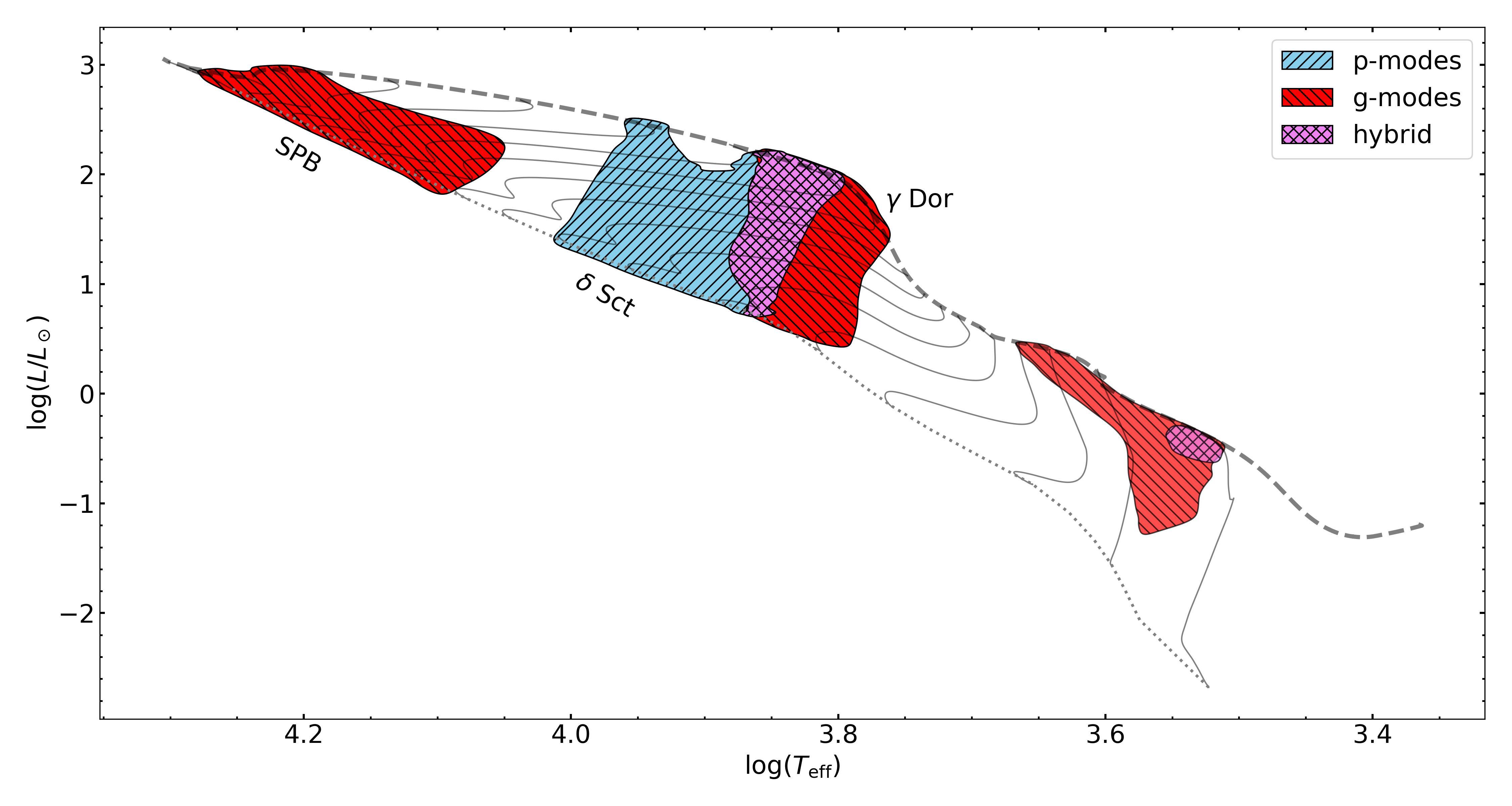}
\end{center}
\caption{Pre-main sequence instability regions in the Hertzsprung-Russell diagram. The colored areas depict the regions for which stellar pulsations are expected to be excited according to the results of \citet{Steindl2021b}.  The dashed grey line shows the evolutionary track of an accreting protostar with time-constant accretion rate. The thin grey lines show the subsequent pre-main sequence evolutionary track evolution and the dotted line indicates the ZAMS.}
\label{fig:instab_strips}
\end{figure}

As of March 2022, seven types of pulsations have been discovered theoretically and observationally in pre-main sequence stars. Sorted from most massive to least massive, these are: Slowly Pulsating B (SPB), $\delta$ Scuti, tidally perturbed, $\gamma$ Doradus, $\delta$ Scuti -- $\gamma$ Doradus hybrid, solar-like, and M type pulsations. Table \ref{tab:preMS_puls:types} provides an overview of their properties and gives approximate current numbers of known objects, and Figure \ref{fig:instab_strips} illustrates the corresponding instability regions.

Overall, the pre-main sequence pulsators have the same pulsation properties as their counterparts in the main sequence and post-main sequence stages. The difference between the evolutionary stages lies in the pattern of excited oscillation frequencies \citep[e.g.,][]{Suran2001,Bouabid2011,Gruber2012} which is another beautiful illustration of the power of asteroseismology. Below we briefly describe the properties of the known types of pre-main sequence pulsators sorted from most massive to least massive.

\textbf{SPB type.} The pulsations in SPB type stars are excited by the heat-engine ($\kappa$) mechamism acting in the ionisation zone of metals \citep{Dziembowski1993}. The pulsation periods lie between about 0.5 and 3 days \citep{Aerts2010}. With masses between $\sim$3 and 7\,\Msun, the pre-main sequence evolution of SPB type stars proceeds relatively fast making them statistically less frequent. As a consequence, SPB pulsators before the onset of hydrogen core burning are observationally harder to find. The expected temperature range for pre-main sequence SPB stars is 11100 to 18700\,K \citep{Steindl2021b}.

\textbf{$\delta$ Scuti type.} The pulsation periods of these intermediate-mass pre-main sequence stars with effective temperatures from 6300 to 10300\,K \citep{Steindl2021b} lie between $\sim$18 minutes and 7 hours \citep{Zwintz2019}. Pre-main sequence $\delta$ Scuti stars show $p$-modes driven by the heat-engine ($\kappa$) mechanism in the ionisation zones of hydrogen and helium \citep{Aerts2010}. This is the group of pre-main sequence pulsators that was discovered first. Because of their pulsation periods, pre-main sequence $\delta$ Scuti stars could easily be detected with ground-based observations obtained only within a few nights. 

\textbf{Tidally perturbed type.} Intermediate-mass $\delta$ Scuti type stars can often be found in binary systems. In some cases, the two components of the binary systems interact leading to strong effects on their structure and evolution \citep[e.g., ][]{DeMarco2017}. If the two components are in a close and eccentric orbit, tidal effects cause self-excited pulsation modes to be perturbed \citep[e.g., ][]{Reyniers2003a,Reyniers2003b}. As of March 2022, only one pre-main sequence star, RS Cha, is known to show tidally perturbed oscillations \citep{Steindl2021a}.

\textbf{$\gamma$ Doradus type.} Pre-main sequence $\gamma$ Doradus stars have early F spectral types. Their expected range in effective temperature lies between 5200 and 7650\,K \citep{Steindl2021b}. 
First theoretical predictions for this type of pulsations in pre-main sequence stars have been conducted by \citet{Bouabid2011} without observational evidence. The first observational detections followed a few years later \citep{Zwintz2013}. The $g$-mode pulsations of pre-main sequence $\gamma$ Doradus stars are excited by the convective flux blocking mechanism \citep{Guzik2000}. The pulsation periods are in the range from 0.3 to 3 days \citep{Aerts2010} and, hence, are quite similar to those in SPB stars. A reliable value for effective temperature is therefore required to identify the type of pulsator as the light curves alone are not sufficient. 

\textbf{$\delta$ Scuti -- $\gamma$ Doradus hybrid type.} Some pre-main sequence pulsators in the A to F range of spectral types can show both $p$- and $g$-modes, hence, $\delta$ Scuti and $\gamma$ Doradus type pulsations. Consequently, this class of objects combines the properties of both classes described above. 

\textbf{Stochastic solar type.} Stochastic solar-like $p$-mode oscillations are predicted to be excited in stars before their arrival on the ZAMS \citep[e.g.,][]{Samadi2005}. Pre-main sequence stars in the mass range of our Sun are mostly very active objects with magnetic fields, spots on their surfaces, and partly still accreting material from circumstellar disks. The light curves obtained for such objects often show regular and irregular variability that is not connected to pulsations. To be able to search for stochastic solar-like oscillations in pre-main sequence stars requires a suitable tool that deals with the high activity which introduces a high background signal \citep{Muellner2021}. Only one candidate is known at the moment \citep{Muellner2021}, but the search continues.

\textbf{K and M type.} This is a recently discovered type of pulsation in pre-main sequence stars that has no known counterpart in the main sequence and post-main sequence phases. \citet{Steindl2021b} found a region of instability for K- and M-type stars which was expected from previous works \citep[e.g.,][]{Baran2011} and presented a first candidate pulsator of this class. The driving mechanism for M-dwarfs is expected to be the $\epsilon$-mechanism \citep[e.g.,][and references therein]{Baran2011} but detailed investigations of the instability regions in \citet{Steindl2021b} have not been performed and are subject of future work.

\section{Challenges}
\label{sec:challenges}
The field of pre-main sequence asteroseismology was met by many challenges throughout its relatively brief history. The initial challenge was taken by \citet{Breger1972} who presented the first evidence for pulsational variability in pre-main sequence stars located in NGC 2264. Since then, due to the advent of space telescopes, the number of known pre-main sequence pulsators has risen above 100. 
In the last decades, lots of challenges regarding pulsations in such young stars have been identified. Many of these have been partly or fully solved, while others remain open until today. The more we start to understand stellar structure and evolution in detail, the more challenges are continuously being created. This sections aims at discussing the currently most important challenges faced by pre-main sequence asteroseismology.

\subsection{Observational challenges}
When observations of young stellar objects shall be conducted, several challenges have to be tackled. These are mainly related to the early evolutionary state of the stars.

\textbf{Activity.} Young protostars are formed in molecular clouds. During their first evolutionary stages, they gain mass by accreting matter from their birth environment. Consequently, young stars can be partially or completely embedded in dense gas and dust, magnetic fields influence how the matter is accreted onto the early star, and the angular momentum gained from the birth process lets the young stellar object spin fast in most cases. All these phenomena can be summarized with the description that young stars show different levels of activity which manifest themselves in our observations. 

The dense circumstellar material can prevent us completely from viewing the young stars in the optical or generates irregular light variations of up to several magnitudes \citep[e.g.,][]{Cody2014}. Slightly less dense material can still be responsible for semi-regular variability \citep[e.g.,][]{Alencar2010}. Searching for millimagnitude pulsations in photometric time series of young stars therefore becomes tricky \citep[e.g.,][]{Zwintz2009}.

The irregular or semi-regular variability originating from the disks has a second challenge for the search and characterization of pulsations: in case the pulsational variability has long periods (i.e., longer than about half a day), the distinction between variability originating from the disk and from the pulsations will be impossible in many cases. The reason is that the irregular variability produces artifacts in the frequency analysis in the low frequency domain where we would also search for the pulsations. Only if the pulsation periods are shorter (i.e., on the order of a few hours and shorter), can they be well distinguished from variability caused by the disk and the artifacts generated during the frequency analysis.

The determination of colors for pre-main sequence stars is also affected by the dense dust that surrounds them: young stars appear much redder than they actually are. Observed colors include the star-disk system and not the star alone. As no general relations for dereddening can be applied for individual young stars with disks (i.e., Herbig Ae/Be stars), the real stellar colors cannot be obtained for embedded objects.

Spectroscopically, the circumstellar matter is visible as very characteristic emission features, for example in the hydrogen lines. Although finding emission lines in the spectra is a good indicator for potentially young stars, in many cases it prevents a reliable calculation of effective temperature and gravitational acceleration which are needed to place the stars into a Kiel diagram.

\begin{figure}[h!]
\begin{center}
\includegraphics[width=\linewidth]{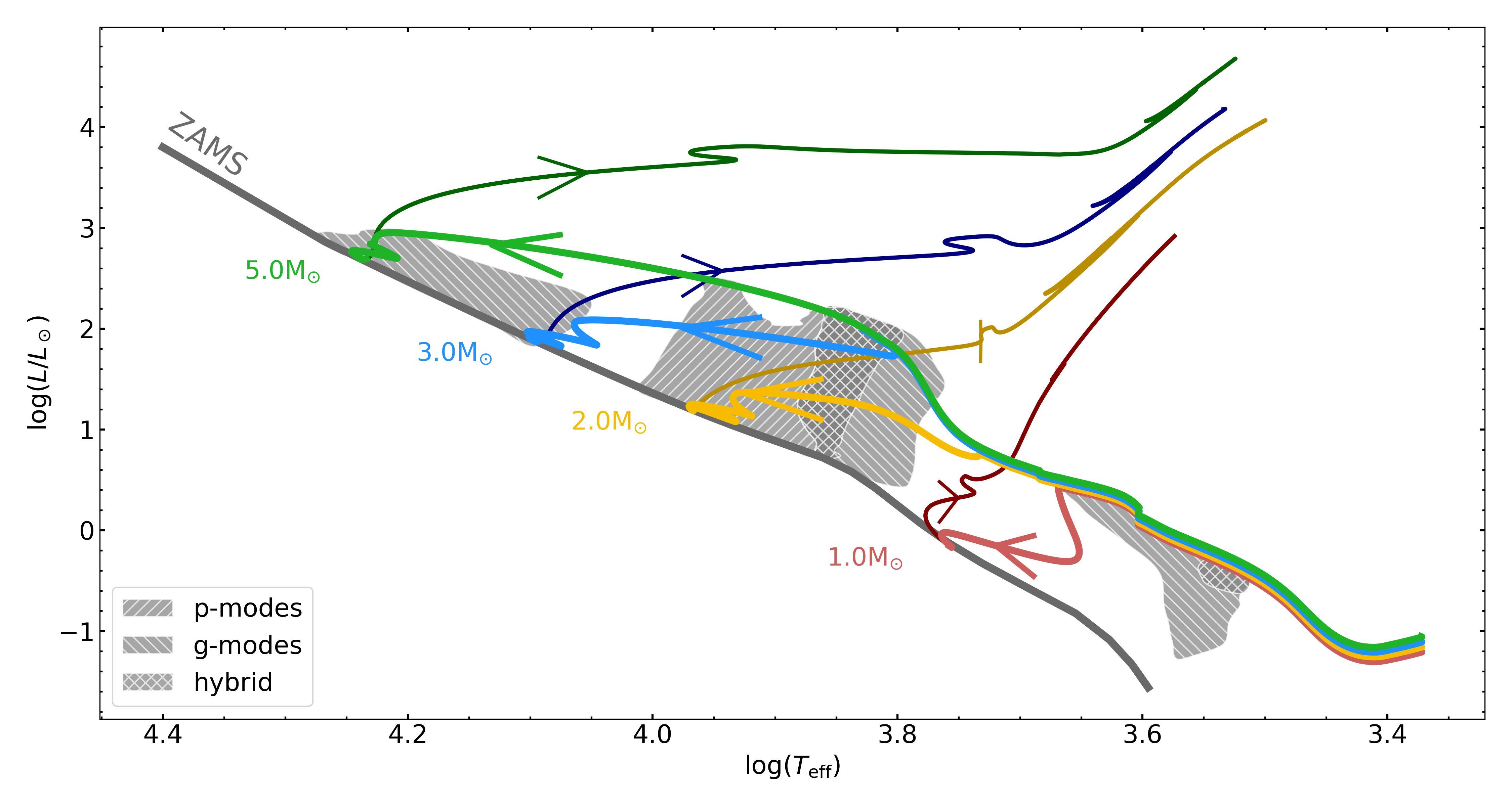}
\end{center}
\caption{Evolutionary tracks for the protostellar seed until the asymptotic giant branch. The colored lines show the evolution of stars with different masses between $1$ and $5\, M_\odot$. The pre-main sequence evolution is shown in brighter colors compared to the main- and post-main sequence evolution. Evolutionary tracks are shifted slightly at the beginning of the evolution for better visibility. The zero age main sequence is shown as grey line and the instability regions (i.e., the same as in Figure \ref{fig:instab_strips}) are shown in grey with hashes according to the legend. }
\label{fig:crossing}
\end{figure}

\textbf{Evolutionary stage.} Taking the atmospheric properties of given stars (i.e., effective temperature, luminosity, and mass) and placing them into a Hertzsprung-Russell diagram does not provide a unique identification of their evolutionary stage which is illustrated in Figure \ref{fig:crossing}. Some observational features related to activity have to be used to collect indications for the young evolutionary stage, and the more of these indicators are present, the better. If stars can be attributed to a star forming region or an open cluster as young as -- say -- ten million years, then this can be considered as excellent evidence for stellar youth. Observational properties such as irregular variability in the photometric time series, infrared and / or ultraviolet excesses, or emission lines in their spectra can point to an early evolutionary stage, but are not unique identifiers because they might also be attributed to quite evolved evolutionary stages. Infrared excesses, for example, can be also found for post-asymptotic giant branch (post-AGB) stars \citep[e.g.,][]{Kamath2014}, and circumstellar material is present in the form of Keplerian disks also around classical Be stars \citep[e.g., ][]{Rivinius2013}.

\textbf{Availability of time-series photometry from space.} Current and former missions have either not targeted young stellar objects or were quite limited in their observations of the early evolutionary phases of stars and planets. 

The currently operational and hugely successful NASA mission TESS \citep{Ricker2015} can reach down to the galactic plane, but the resulting light curves often suffer from high contamination. The reason is that the CCD pixels are relatively large (i.e., 21 arcseconds per pixel). Consequently, TESS observations avoid to observe deep in the galactic plane. 

The NASA mission Kepler \citep{Borucki2010} observed a single field high above the galactic plane on purpose to avoid star forming regions and any resulting contamination. The Kepler K2 \citep{Gilliland2010} mission provided some data for young stars and star forming regions in four of 19 campaigns (i.e., campaigns numbers 2, 9, 13, and 15) illustrating the potential of space observations for this research field.

The BRITE-Constellation nano-satellite mission \citep{Weiss2014} targets only the brightest stars on the sky, limiting the observations of young stars and planets (which are typically fainter by several orders of magnitude) dramatically. 

The earlier satellite missions CoRoT \citep{Auvergne2009} and MOST \citep{Walker2003} allowed for observations of the youngest objects in the galaxy through dedicated short (i.e., between 10 days and 5-6 weeks) observing runs, for example on the young cluster NGC 2264 (MOST \& CoRoT) or on individual young stellar objects such as HD 142666, HD 37806, or TW Hya.

ESA’s future mission PLATO \citep[planned launch in 2026; ][]{Rauer2014} is scheduled to observe two selected fields for two years each: both fields will not reach down to the galactic plane, hence not be able to target the youngest regions in the Milky Way. Additionally, PLATO’s pixel size of 18\,$\mu$m $\times$ 18\,$\mu$m yields a plate scale of about 26.5 arcseconds per pixel which is even higher than TESS’s plate scale. Therefore, observations of star forming regions and young clusters with high object densities will be problematic for PLATO due to high percentages of contamination.

The current maximum time bases for continuous photometric observations of pre-main sequence pulsating stars are $\sim$80 days from Kepler K2 and slightly more than 100 days from TESS \citep{Steindl2021b}. Therefore, pre-main sequence asteroseismology has the challenge to work with way more limited observational material than most of the other fields in asteroseismology.

\subsection{Theoretical challenges}

Many ingredients are needed to properly describe the earliest phases of stellar evolution since many physical processes are active during that time span. In terms of complexity of the pre-main sequence evolution, the discussion in the introduction only scratches the surface of the challenges we are faced with to create theoretical models of such stars. Stellar rotation, magnetic fields, and star-disk interaction are just a few examples of the physical ingredients, in addition to mass accretion, that need to be kept in mind. All of the above will generally be different for every object. Hence, there might not be a single other phase of stellar evolution in which the spectroscopic parameters and internal structure vary as much on a case by case level as during the pre-main sequence evolution.

\textbf{Stellar rotation.} When stars are born in the collapse of a molecular cloud, they obtain angular momentum. Throughout the accretion phase, in which material from the surrounding disk deposited onto stellar surface, the system is expected to be disk-locked \cite{Bouvier1997}. That is, the disk and the star co-rotate until the former is dissolved or its influence on the star becomes minor. The mechanism of disk-locking, however, provides many open questions for the implementation of pre-main sequence models: How long does the disk-locking phase last? What is the distribution of rotation periods and how is it produced? Does the disk lock only the stellar atmosphere or is the whole star co-rotating? If the former, how is the angular momentum distributed in the stellar atmosphere and what is the mechanism of the angular momentum transport? If the latter, what mechanism fixes the rotation rate throughout the star? Some of these questions linger to even later phases of the pre-main sequence stage. After the disk has dissolved, angular momentum throughout the star will evolve according to a not yet fully explained mechanism. Including angular momentum into the current description of stellar evolution models remains an open question with lots of impact on the pulsational characteristics of stars: For gravity mode pulsators, the period spacings are tilted according to the angular momentum of the core while the frequencies of pressure modes are split with respect to the angular velocity and the azimuthal order \citep{Aerts2010}. The Coriolis force in rotating stars gives rise to a new family of pulsation modes, the Rossby modes. The latter have so far not been detected in any pre-main sequence object. 

\textbf{Magnetic fields.} As is common in the theory of stellar structure and evolution, many of the theoretical challenges of pre-main sequence asteroseismology are intertwined. Magnetic fields, for example, are expected to play a major role in the rotational evolution of stars. As such, they are expected to dominate the angular momentum transport in radiative zones, albeit not efficient enough to explaining observations \citep[i.e.][]{Fuller2014}. Magnetic braking seems to be the dominant mechanism for angular momentum loss in more evolved stars \citep[i.e.][]{Matt2015}. For pre-main sequence stars, magnetic fields are expected to be an important ingredient for disk-locking \citep{Barnes2001}. As the latter already implies, magnetic fields also carry implications on the mass accretion mechanism and hence the accretion rates themselves \citep[i.e.][]{Bouvier2007}. In addition, magnetic fields directly effect the internal structure of stars including the mode cavities and leave a measurable imprint on the pulsation frequencies \citep{Prat2019}. The interaction between magnetic fields and pulsation can lead to a suppression of the latter, resulting in a change in mode amplitudes \citep[i.e.][and references therein]{Lecoanet2022}. Magnetic fields with strengths of multiple kG have been found in pre-main sequence stars \citep{Lavail2017} while their consequences for pre-main sequence asteroseismology have not been explored. 

\textbf{Mass accretion rates.} 
The atmospheric parameters of pre-main sequence stars \citep{Steindl2021b, Steindl2022b} as well as their internal structure \citep[see][and the discussion in the introduction to this article]{Steindl2022a} are dependent on the characteristics of the accretion process. While time-dependent mass accretion rates are, although limited in amount, readily available from 2-dimensional simulations of the disk \citep[e.g.][]{Vorobyov2015, Jensen2018,Elbakyan2019}, many other free parameters need to be set in the calculation of stellar structure models. Most noteworthy, we lack an intrinsic description of the mechanism that describes the energy flow of the accreted material. How much energy is added to the star? How much is radiated away? Where is the energy deposited? At the current stage, we have to manually set many parameters corresponding to different assumptions. For further progress in this field it is inevitable to investigate the detailed physics of the accretion processes in more detail.
Additional effects complicate the calculation of the equilibrium stellar structure. The properties of the material transferred from the accretion disk to the star are expected to be dependent on the accretion rate itself. For example, the metallicity should follow the relation $Z_{\rm acc} = \frac{\dot{M}_d}{\dot{M}_{\rm acc}}$ \citep{Kunitomo2021}, where $\dot{M}_d$ is the flux of the gas and $\dot{M}_{\rm acc}$ is the mass accretion rate. Many of the to-date calculated mass accretion histories cannot deliver the needed information of $\dot{M}_d$. However, recent studies provide this information \citep[see e.g.][]{Elbakyan2020} such that the inclusion of effects from condensing material will be possible in the near future. Most probably, however, the inclusion of these effects will further push the software instrument Modules for Experiments in Stellar Astrophysics (\textit{MESA}) \citep{paxton2011, paxton2013, paxton2015, paxton2018, paxton2019}. \textit{MESA} was never designed to perform such calculations and, albeit providing us with an indispensable and vital tool, repeatedly runs intro convergence issues with strong time-dependent mass accretion rates during the early phases of the pre-main sequence evolution. Strong efforts will need to go into \textit{MESA}-related problems which is a time consuming work.
However, we are not concerned that progress in this regard will have to wait long since the core \textit{MESA} team is very helpful in any regards of their community focused tool. 

\textbf{The issue of controlled grid studies.} 
The simple fact that each and every pre-main sequence star has its very own time-dependent accretion rate history and, hence, a very different (and often times chaotic) evolution in the Hertzsprung-Russell diagram \citep{Steindl2022a} complicates the calculation of controlled grids. With the inclusion of disk-mediated accretion rates, the days of (almost) parallel evolutionary tracks are gone which complicates almost every theoretical study. Incorporating assumptions similar to those in the work of \citet{Steindl2021b}, namely that each star follows the same accretion track (with constant mass accretion rate) simplifies such studies, but at the cost of completely disregarding the different evolutionary paths. Quasi-random grids, similar to \citep{Steindl2022b} are in general to be preferred, but the exact values of parameters at a given location in the Hertzsprung-Russell diagram are then not uniquely defined by one evolutionary track. Albeit disk-mediated mass accretion rates are available in a limited amount, the calculation of thousands (as we would wish for in such studies) remain challenging due to the needed computational time. 

\textbf{Pre-main sequence asteroseismology beyond intermediate mass stars.}
Among known pre-main sequence pulsators, $\delta$ Scuti stars significantly outnumber both $\gamma$ Doradus and SPB stars \citet{Steindl2021b}. While it is reasonably simple to verify the pre-main sequence status for $\delta$ Scuti and $\gamma$ Doradus stars, such a verification is much more complicated for the more massive SPB stars. Owing to the fast evolution towards the main sequence, it remains a matter of debate at which mass range will it be still possible to observe stars in their pre-main sequence stage. This, of course, will again be dependent on their evolutionary path from the protostellar stage to the ZAMS. This calls in the need for dedicated calculations with disk-mediated accretion rates that end in higher mass stars \citep{Steindl2022b}. This will not only be helpful in regard to SPB stars, but should provide many insights in asteroseismology of $\beta$ Cephei stars with even higher mass as well. 
In the low mass regime, theoretical models suggest an instability region for K- and M-type stars \citep[i.e.][]{rodriguez2019, Steindl2021b} and a first candidate for such pulsation has been presented by \citet{Steindl2021b}. According to the theoretical models, many radial orders of g-modes seem to be excited \citep{Steindl2021b}. This instability region needs to be further explored with improved theoretical models for which an important step is to further decrease the mass of the initial stellar seeds which is usually taken to be $\sim10\,M_{jupiter}$ \citep{Steindl2021b, Steindl2022a, Steindl2022b}.

\section{STRETTO}
\label{STRETTO}
STRETTO (Early STaRs and planEt evoluTion in Two cOlors) is an innovative project idea that aims to provide a micro-satellite for astronomy from space with the main goal to study early stellar and planetary evolution.

\textbf{Science goals.} STRETTO aims to investigate young stars and planets in star forming regions as well as the youngest open clusters with the goal to address their early evolution.
The STRETTO space telescope will be able to study the strength and properties of stellar activity and the amount of rotation present in early stars and their influence on planet formation and evolution. STRETTO will search for signs for the formation of planets and the presence of planets around member stars of young open clusters and star forming regions. The photometric time series obtained by STRETTO will enable studies of the effects of accretion on young stellar objects, of eclipsing binary and multiple systems in their early evolutionary stages, and the interior structures of young stars using asteroseismology. The expected precision of STRETTO will let us investigate the properties of ring systems around exoplanets, other circumplanetary material and the existence of smaller bodies (e.g., exomoons or exocomets) around young stellar objects. Also, the properties of young open clusters and star forming regions as larger-scale objects in our universe can be investigated with such a mission.

Together with complementary ground-based observations, STRETTO science will allow to improve the input physics for the early phases of stellar and exoplanetary evolution, provide a time-dependent map of rotation and chemical composition from stellar birth to the onset of hydrogen-core burning, determine a complete picture of the angular momentum transport of young stars from the
interior to the atmosphere, provide more reliable ages for the youngest stellar and exoplanetary objects, investigate the connection between magnetic fields and variability of stars in their early evolutionary stages, and understand the interaction of the young circumstellar environment with the star, including exoplanets, exomoons, and exocomets.

\textbf{Instrumental design.} STRETTO will carry two 8-cm telescopes each with a 1.5 x 1.5 square degree field of view and a spatial resolution of 3 – 5 arcseconds per pixel. Each telescope will have a dedicated filter: one in the optical, the other at infrared wavelengths. From a low-Earth orbit, STRETTO will be able to monitor the young stars and planets for about half a year continuously, providing the necessary long time-bases for the analysis of the objects’ different types of variability. STRETTO will be able to take photometric time series measurements of stars in the magnitude range from about 6 to 16 mag (V) in two colors using alternating exposure times in the range from 1 to 60 seconds.
The goal is to utilize a commercially available microsatellite platform (mass: 50-70kg and power: 60- 80W) which shall host two digital camera systems as payloads, one for each passband. A low earth polar orbit in the height range of 600 – 900 km will be suitable to conduct the scientific observations. The baseline for communication will be one ground station in Europe.

\textbf{Potential of STRETTO.}
The scientific potential of monitoring young stars and planets photometrically from space with STRETTO lies in yielding a first clear picture how stars and planets pass through their earliest evolutionary phases. 

\textbf{Status of the project.} Presently, a small consortium consisting of researchers and engineers from Austria, Canada, France, the Netherlands, and Poland is trying to acquire funding for a concept study. If you are interested to learn more about STRETTO and the people involved, please contact the first author of this article.

\section{Prospects and importance}
\label{sec:prospects}

One might think that our theoretical understanding of stellar evolution is well established with only a minor need for further research. But this is a misconception as there are many physical processes that are either not well-understood (e.g., the impact of accretion on the complete evolution of stars) or not taken into account properly in our theoretical models (e.g., convection, rotation or magnetic fields). The physical effects occurring and defining the earliest evolutionary phases of stars must have an impact on their complete further evolution. It would be physically unlikely that the stars' formation histories do not play a role in their later stages. 

One of the biggest questions in this respect is how large the impact of the processes acting in the youngest stellar objects is and how long the pre-main sequence history of stars persists up to later stages. This is one of the questions pre-main sequence asteroseismology can and does address. By adding processes such as accretion to theoretical models of pre-main sequence stars and coupling those to models of pulsational instability lets us investigate the resulting changes in the interior structures of stars. 

Pre-main sequence asteroseismology should be able to test the very early evolutionary phases as well. By studying objects with ages of a few million years, we can measure the imprint of the star formation process, thereby shedding light on the many free parameters connected to the accretion physics. The amazing prospect of gathering observational information about the internal structure of stars (rotation rate, chemical mixing profiles, etc.) opens the door to exciting constraints for theoretical models. As of today, the earliest evolutionary phases of stars are often treated as a black box using crude approximations as in the classical model. The resulting stellar structure and atmospheric parameters, however, are used in many different fields to motivate for example the existence of magnetic fields or the evaporation of exoplanet atmospheres. Only dedicated asteroseismic studies of the youngest objects we can possible find, can provide us with the important ingredients to study these processes with the accuracy they deserve. 

Chemical composition plays an important role in stellar structure and evolution due to the sensitivity of opacities on the atomic spectra and absorption features of the elements making up the star. Most stars pulsate exactly because of the behaviour of the opacities in relation to perturbations (e.g., the heat engine mechanism). Also, the location of the computed evolutionary tracks for stars at all ages depends on the metallicity, Z \citep[e.g.,][]{Montalban2004}.
Presently, we do not understand how the chemical evolution proceeds between stellar birth and the onset of hydrogen core burning upon arrival on the ZAMS. For example, in $\sim$10\% of main sequence stars of spectral type B to F chemical peculiarities are found \citep[e.g.;][]{Preston1974}, but it is unclear when these anomalies are formed. The first few detailed analyses of the atmospheric chemical abundances of pulsating pre-main sequence stars have revealed basically solar or slightly solar chemical composition with two exceptions: (i) stars with masses smaller than $\sim$1.5\,\Msun have not burnt the primordial Lithium completely and, hence, show an overabundance compared to the Sun \citep{Zwintz2013}; (ii) in the high-resolution spectra of intermediate-mass pre-main sequence pulsators Barium shows a significant overabundance which cannot be fully explained yet \citep[e.g.,][]{Zwintz2013}. In the future, high-resolution spectroscopic observations of a statistically large enough sample of pre-main sequence stars should be used to generate a time dependent map of the chemical evolution in the early stages of the lives of stars. 

Asteroseismology has successfully revealed the interior chemical structure of stars: it allows measuring the percentage of hydrogen in the cores of main sequence stars \citep[e.g.,][]{Moravveji2015} or detecting chemical gradients in g-mode period spacings \citep[e.g.,][]{Miglio2008,Bouabid2013}. Consequently, pre-main sequence asteroseismology has the potential to probe the interior chemical evolution of stars in the earliest phases of their evolution. Possible topics in this context would be to investigate the influence of the accretion history on the chemical evolution of stars and how long it persists, if observed chemical inhomogeneities on the stellar surfaces extend into the interiors or not, and if stellar pulsations let us deduce, for example, the amount of Deuterium in the earliest stars. But such investigations require an improvement in our theoretical models and dedicated instruments providing the high-accuracy data (both photometrically and spectroscopically) for pre-main sequence stars. 

As the excited pulsation frequencies in pre-main sequence stars are different to those in the post-main sequence stages due to the different inner structures \citep[e.g.,][]{Suran2001,Bouabid2011}, it is obvious that stellar pulsations can be used to distinguish the evolutionary stages of stars \citep{Guenther2007}. Even within the pre-main sequence stages, the pulsational properties of stars change following a relation that is not present for the same type of pulsators in later phases: the youngest objects pulsate slower than stars close to the onset of hydrogen core burning \citep[i.e., the ZAMS, ][]{Zwintz2014}. Therefore the next logical step is to use the pulsation properties of pre-main sequence stars as an age indicator for stellar astrophysics. In the earliest evolutionary phases, it is difficult to determine precise ages based on our currently available methods. The ages of the youngest open clusters, for example, are typically given with errors of 50 to 100\,\%. One of the important prospects of pre-main sequence asteroseismology is therefore to provide accurate (relative) ages for young stellar objects - similar to the percentage of hydrogen in the core, $X_c$, that is determined for the main sequence stages from asteroseismology. This is especially important, as different age indicators that are very useful for the study of older clusters, often fail to improve the accuracy of the age determination of young open clusters. Gyrochronology, for example, can provide fantastic constraints on the age of open clusters, but the usage for pre-main sequence stars is very limited. Measurements of the surface rotation periods of young stars are available, but the only way to reproduce them theoretically is to force a specific distribution of initial rotation periods. One of the major issues in this regard is the effect of the protostellar disk during the accretion phase. By coupling stellar evolution codes with a designated disk-evolution, we are hopeful to improve the models in this regard. Once we have an accurate picture of the rotation of pre-main sequence stars, we can explore the effects of the stars' rotation on their pulsational properties in much more detail.

Asteroseismology of pre-main sequence stars is needed to address the question why intermediate-mass stars on the main sequence tend to show rigid rotation independent of their core rotation rates \citep{Aerts2017}. Strong coupling between the stellar core and the envelope seems to occur for stars on the main sequence and in later evolutionary phases. With pre-main sequence asteroseismology we will be able to investigate at what earlier point in stellar evolution this strong coupling starts. By measuring nearly-equidistant period spacings we can deduce near-core rotation rates for pre-main sequence g-mode pulsators -- as it is already successfully done for stars in later evolutionary stages. First steps in these investigations have been undertaken, but for a complete picture of the angular momentum transport in young stars, longer photometric time-series with highest precision obtained from space are required. The observational material available for pre-main sequence g-mode pulsators now is insufficient to conduct more detailed studies. 
As a consequence, the idea of the micro-satellite STRETTO (see Section \ref{STRETTO}) dedicated to young star- and planet-forming regions emerged a couple of years ago and will hopefully be realized in the near future.

\section*{Conflict of Interest Statement}

The authors declare that the research was conducted in the absence of any commercial or financial relationships that could be construed as a potential conflict of interest.

\section*{Author Contributions}
K. Zwintz and T. Steindl shared the work on this article and contributed in equal amounts to it.


\section*{Funding}
K. Zwintz and T. Steindl are funded by the University of Innsbruck and are grateful for the support they are receiving.

\section*{Acknowledgments}
We thank Matthew Kenworthy from the University of Leiden (NL), Gregg Wade from the Royal Military College (CAN) and Rainer Kuschnig from the University of Graz (AT) for their collaboration on the satellite project STRETTO. We are grateful to Eduard Vorobyov from the University of Vienna (AT) for the contribution of the time-dependent accretion rates.




\bibliographystyle{frontiersinSCNS_ENG_HUMS} 
\bibliography{test}




\begin{table}
 \begin{scriptsize}

    \caption{Types of known pre-main sequence pulsators and their properties (as of March 2022).}
    \label{tab:preMS_puls:types}
            \smallskip
\begin{tabular*}{\linewidth}{lllcr}
        \hline
        \noalign{\smallskip}
        Pulsation type   &  Mass range & Periods & Known objects & References\\
        \noalign{\smallskip}
        \hline
        \noalign{\smallskip}
        SPB & 3.0 -- 7.0\,\Msun  & 0.8\,d -- 3\,d & 18 & [1], [2] \\
        $\delta$ Scuti & 1.5 -- 2.5\,\Msun  & 18\,min -- 8\,h & $>$100 & [3]\\
        Tidally perturbed  & 1.8 -- 1.9\,\Msun & 48\,min -- 3.3\,h & 1 & [4] \\
        $\gamma$ Doradus &  1.4 -- 1.8\,\Msun & 0.3\,d -- 3\,d & 8 & [5], [6]\\
        $\delta$ Scuti - $\gamma$ Doradus hybrids & 1.4 - 2.2\,\Msun  & 18\,min -- 8\,h \& 0.3\,d -- 3\,d & 4 & [7] \\
        Solar-like & $\sim$\,1\Msun  & $\sim$1.15\,h$^1$ & 1$^{\star}$ & [8] \\
        M-stars & $\sim$\,0.15\Msun  & 0.5\,d - 5\,d & 1$^{\star}$ &[6]\\
    \noalign{\smallskip}
    \hline
    \end{tabular*}    
    \small{References: [1] \citet{Gruber2012}, [2] \citet{Zwintz2017}, [3] multiple papers, e.g., \citet{Zwintz2008,Zwintz2014,Mellon2019,Steindl2021b}, [4] \citet{Steindl2021a}, [5] \citet{Zwintz2013}, [6] \citet{Steindl2021b}, [7] \citet{Ripepi2010}, [8] \citet{Muellner2021}. \\
    $^1$ ... value for $\nu_{\rm max}$; \\
    $\star$ ... Only candidate stars detected so far.}
 \end{scriptsize}
\end{table}

\end{document}